\documentstyle[epsf]{elsart}

\newcommand{\bbox}[1]{\mbox{\boldmath$#1$}}

\makeatletter
\def\lsim{\mathrel{\mathpalette\gl@align<}}
\def\gsim{\mathrel{\mathpalette\gl@align>}}
\def\gl@align#1#2{\lower.6ex\vbox{\baselineskip\z@skip\lineskip\z@
    \ialign{$\m@th#1\hfil##\hfil$\crcr#2\crcr\sim\crcr}}}
\makeatother

\begin{document}

\begin{frontmatter}

\title{Relativistic and Non-Relativistic Mean Field Investigation
of the Superdeformed Bands in $^{62}$Zn}

\author[Mado1,Mado2]{Hideki Madokoro\thanksref{Email1}} and
\author[Matsu]{Masayuki Matsuzaki\thanksref{Email2}}
\address[Mado1]{Department of Physics, Kyushu University, Fukuoka 812-8581,
Japan}
\address[Mado2]{Cyclotron Laboratory, The Institute of Physical and Chemical
Research(RIKEN), Wako, Saitama 351-0198, Japan}
\address[Matsu]{Department of Physics, Fukuoka University of Education,
Munakata, Fukuoka 811-4192, Japan}
\thanks[Email1]{Electronic address: madokoro@postman.riken.go.jp}
\thanks[Email2]{Electronic address: matsuza@fukuoka-edu.ac.jp}

\begin{abstract}
Following the discovery of the superdeformed(SD) band in  $^{62}$Zn,
we calculate several low-lying SD bands in $^{62}$Zn using
Relativistic Mean Field and Skyrme-Hartree-Fock models.  Both models
predict similar results, but still we can see some qualitative
differences in the results of these two models, which are coming from
the difference of the detail of single-particle levels.
\end{abstract}

\end{frontmatter}

\newpage

Recent development of the experimental techniques enables us to study
the nuclei at very high angular momentum.  The study of the superdeformed(SD)
bands is one of the most interesting topics in such studies.  The first
experimental discovery was done for $^{152}$Dy in 1986\cite{ref:Tw86}. Since
then, a large number of the SD bands have been observed in the $A\sim$150,
130, 190 and 80 mass regions.  These SD states are generated on the second
minima in the potential energy surface which are connected with the deformed
shell gaps such as $N, Z\sim$44, 64, 86 and 116. Based on the theoretical
calculation indicating that there exist the $N,Z\sim$30 shell gaps,
the SD bands in the $A\sim$60 mass region were also predicted\cite{ref:Ra90}.
In spite of much experimental effort, however, such SD states had not been
observed due to some experimental difficulties.  In 1997, a cascade of six
$\gamma$ rays forming a new band was observed in $^{62}$Zn\cite{ref:Sv97},
which was the first experimental discovery of the SD bands in this mass
region.  The extracted $\beta_{2}$ value is 0.45$^{+0.10}_{-0.07}$, and
the SD states seem to become yrast at $I\gsim$24 in comparison with a
theoretical calculation\cite{ref:Sv97}.  This spin value corresponds to the
rotational frequency $\Omega\gsim$1.3 MeV which is the highest frequency
among those observed so far.  Besides, there are several interesting features
in this mass region, for example, that the valence neutrons and protons
occupy the same orbits in contrast with those in other mass regions and the
residual neutron-proton pairing at high spin may be observed.  Another
example is the decay of these SD shapes via proton emission due to a very
low Coulomb barrier.  Such decay of the well-deformed high spin states has
been already observed in $^{58}$Cu\cite{ref:Ru98_1}.  For this $A\sim$60
mass region, more and more new experimental data will become available in
near future and so many theoretical investigations will be done from now on,
using both the semi-phenomenological models such as cranked Nilsson
(or Woods-Saxon)-Strutinski model and fully microscopic ones, for example,
cranked Hartree-Fock model with density dependent effective interactions
and Relativistic Mean Field(RMF) model.

RMF model is now considered as a reliable method likely to the
non-relativistic Hartree-Fock model to describe various properties of finite
nuclei, not only $\beta$-stable but also $\beta$-unstable ones.  Since the 
work of Walecka and his colla\-bo\-ra\-tors\cite{ref:ChWa74,ref:Wa74,ref:Ch77},
RMF model has been  applied to nuclear matter and the ground states of finite
nuclei as well as the scattering data with great successes.  Some groups also
tried to apply this model to the excited states in finite nuclei.  In 1989,
Munich group made a first attempt to describe the properties of rotating
nuclei\cite{ref:KoRi89,ref:KoRi90}.  They applied this model mainly to the
SD bands in the $A\sim$150, 80 and 190 mass regions
\cite{ref:KoeRi93,ref:AfKoeRi96_1,ref:AfKoeRi96_2,ref:Koe96_1}. 
The RMF calculations of rotating nuclei are, however, limited to the nuclei
in these mass regions up to now.  Because RMF model is on its way of
refinement, it is surely important to apply this model to various nuclear 
phenomena as wide as possible, and to check its applicability further.
Therefore, in this letter, we apply RMF model to the description of the SD 
bands in $^{62}$Zn including the newly discovered one, and examine how well
this model can describe the very high spin states in neutron-deficient
unstable nuclei.  This is the first RMF calculation of the SD bands
in this mass region.  For comparison, we also calculate the SD bands in
$^{62}$Zn using Skyrme-Hartree-Fock(SHF) model.

The starting point of RMF model is the following Lagrangian which contains
the nucleon and several kinds of meson fields such as $\sigma$-, $\omega$-
and $\rho$-mesons, together with the photon fields(denoted by $A$)
mediating the Coulomb interaction,
\begin{displaymath}
  {\cal L}={\cal L}_{N}+{\cal L}_{\sigma}+{\cal L}_{\omega}+{\cal L}_{\rho}
  +{\cal L}_{A}+{\cal L}_{\rm int}+{\cal L}_{\rm NL},
\end{displaymath}
\begin{eqnarray}
  {\cal L}_{N} & = & \overline{\psi}
    (i\gamma^{\alpha}\partial_{\alpha}-M)\psi, \nonumber \\
  {\cal L}_{\sigma} & = & \frac{1}{2}(\partial_{\alpha}\sigma)
    (\partial^{\alpha}\sigma)-\frac{1}{2}m_{\sigma}^{2}\sigma^{2}, \nonumber \\
  {\cal L}_{\omega} & = &
    -\frac{1}{4}\Omega_{\alpha\beta}\Omega^{\alpha\beta}
    +\frac{1}{2}m_{\omega}^{2}\omega_{\alpha}\omega^{\alpha}, \nonumber \\
  {\cal L}_{\rho} & = &
    -\frac{1}{4}\bbox{R}_{\alpha\beta}\cdot\bbox{R}^{\alpha\beta}
    +\frac{1}{2}m_{\rho}^{2}\bbox{\rho}_{\alpha}\cdot\bbox{\rho}^{\alpha},
    \nonumber \\
  {\cal L}_{A} & = &
    -\frac{1}{4}F_{\alpha\beta}F^{\alpha\beta}, \nonumber
\end{eqnarray}

\noindent
where

\begin{eqnarray}
\Omega_{\alpha\beta} & = & \partial_{\alpha}\omega_{\beta}-
  \partial_{\beta}\omega_{\alpha}, \nonumber \\
\bbox{R}_{\alpha\beta} & = & \partial_{\alpha}\bbox{\rho}_{\beta}-
  \partial_{\beta}\bbox{\rho}_{\alpha}-g_{\rho}\bbox{\rho}_{\alpha}\times
  \bbox{\rho}_{\beta}, \nonumber \\
F_{\alpha\beta} & = & \partial_{\alpha}A_{\beta}-\partial_{\beta}A_{\alpha},
  \nonumber
\end{eqnarray}

\noindent
are the field strength tensors. ${\cal L}_{\rm int}$ is the interaction
part between nucleons and mesons,

\begin{displaymath}
  \begin{array}{ccl}
  {\cal L}_{\rm int} & = & g_{\sigma}\overline{\psi}\psi\sigma
    -g_{\omega}\overline{\psi}\gamma^{\alpha}\psi\omega_{\alpha} \\
  & & -g_{\rho}\overline{\psi}\gamma^{\alpha}\bbox{\tau}\psi\cdot
    \bbox{\rho}_{\alpha}-e\overline{\psi}\gamma^{\alpha}
    \frac{\displaystyle 1-\tau_{3}}{\displaystyle 2}\psi A_{\alpha}.
  \end{array}
\end{displaymath}

\noindent
In the standard applications, the non-linear self interactions among the
$\sigma$-mesons,

\begin{displaymath}
  {\cal L}_{\rm NL}=\frac{1}{3}g_{2}\sigma^{3}-\frac{1}{4}g_{3}\sigma^{4},
\end{displaymath}

\noindent
are also included, which are crucial for the realistic description of
deformed nuclei.  Applying the variational principle to the Lagrangian
gives the equations of motion.  Within the mean field approximation, these
are the Dirac equation for single nucleon fields $\psi_{i}$ and the
Klein-Gordon equations for the classical meson and photon fields.
After solving these equations, we can calculate various properties of
finite nuclei.

For the application to rotating nuclei within the cranking assumption, it is
necessary at first to write the Lagrangian in the uniformly rotating frame
which rotates around the $x$-axis with a constant angular velocity $\Omega$,
from which the equations of motion in this rotating frame can be obtained.
Because the rotating frame is not an inertial one, a fully covariant
formulation is desirable and we accomplished this using the technique of
general relativity known as tetrad formalism\cite{ref:We72,ref:BiDa82}.
The procedure is as follows: First, according to tetrad formalism, we can 
write the Lagrangian in the non-inertial frame represented by the metric 
tensor $g_{\mu\nu}(x)$.  Then the variational principle gives the equations of 
motion in this non-inertial frame.  Finally, substituting the metric tensor of 
the uniformly rotating frame,

\begin{displaymath}
  g_{\mu\nu}(x)=
  \left(
    \begin{array}{cccc}
      1-\Omega^{2}(y^{2}+z^{2}) & 0 & \Omega z & -\Omega y \\
      0 & -1 & 0 & 0 \\
      \Omega z & 0 & -1 & 0 \\
      -\Omega y & 0 & 0 & -1
    \end{array}
  \right),
\end{displaymath}

\noindent
leads to the desired equations of motion.  The resulting equations are

\begin{eqnarray}
  \left\{
    \bbox{\alpha}\cdot(\frac{1}{i}\bbox{\nabla}-g_{\omega}
    \bbox{\omega}(x))+\beta(M-g_{\sigma}
    \sigma(x))\right. & & \nonumber \\
    \left. +g_{\omega}\omega^{0}(x)
    -\Omega(L_{x}+\Sigma_{x})\right\}
  \psi_{i}(x) & = & \epsilon_{i}\psi_{i}(x), \nonumber \\
  \left\{
    -\bbox{\nabla}^{2}+m_{\sigma}^{2}-\Omega^{2}L_{x}^{2}
  \right\}
  \sigma(x) & = & g_{\sigma}\rho_{s}(x), \nonumber \\
  \left\{
    -\bbox{\nabla}^{2}+m_{\omega}^{2}-\Omega^{2}L_{x}^{2}
  \right\}
  \omega^{0}(x)
  & = & g_{\omega}\rho_{v}(x), \nonumber \\
  \left\{
    -\bbox{\nabla}^{2}+m_{\omega}^{2}-\Omega^{2}(L_{x}+S_{x})^{2}
  \right\}
  \bbox{\omega}(x) & = & g_{\omega}
  \bbox{j}_{v}(x), \nonumber
\end{eqnarray}

\noindent
where the $\rho$-meson and photon fields are omitted for simplicity
although they are included in the numerical calculation. These equations
are the same as those of Munich group.  For detail, see ref.\cite{ref:MaMa97}. 

These equations are solved by the standard iterative diagonalization method
using the three-dimensional harmonic oscillator eigenfunctions.  Because
our method of numerical calculation is essentially the same as that of Munich 
group\cite{ref:Koe96_2} and its details are shown in ref.\cite{ref:GaRiTh90},
we here give only the model parameters used.  The cutoff parameters for the
nucleon and the meson fields are taken to be $N_{\rm F}$=8 and  $N_{\rm B}$=10,
respectively.  The parameter set called NL-SH is adopted which is expected to
give a better description of $\beta$-unstable nuclei\cite{ref:ShNaRi93}
than others. Of course the dependence on the parameter set used should be
examined in detail, which will be discussed in a near future work.

\begin{figure}[h]
  \epsfxsize=8cm
  \epsfbox{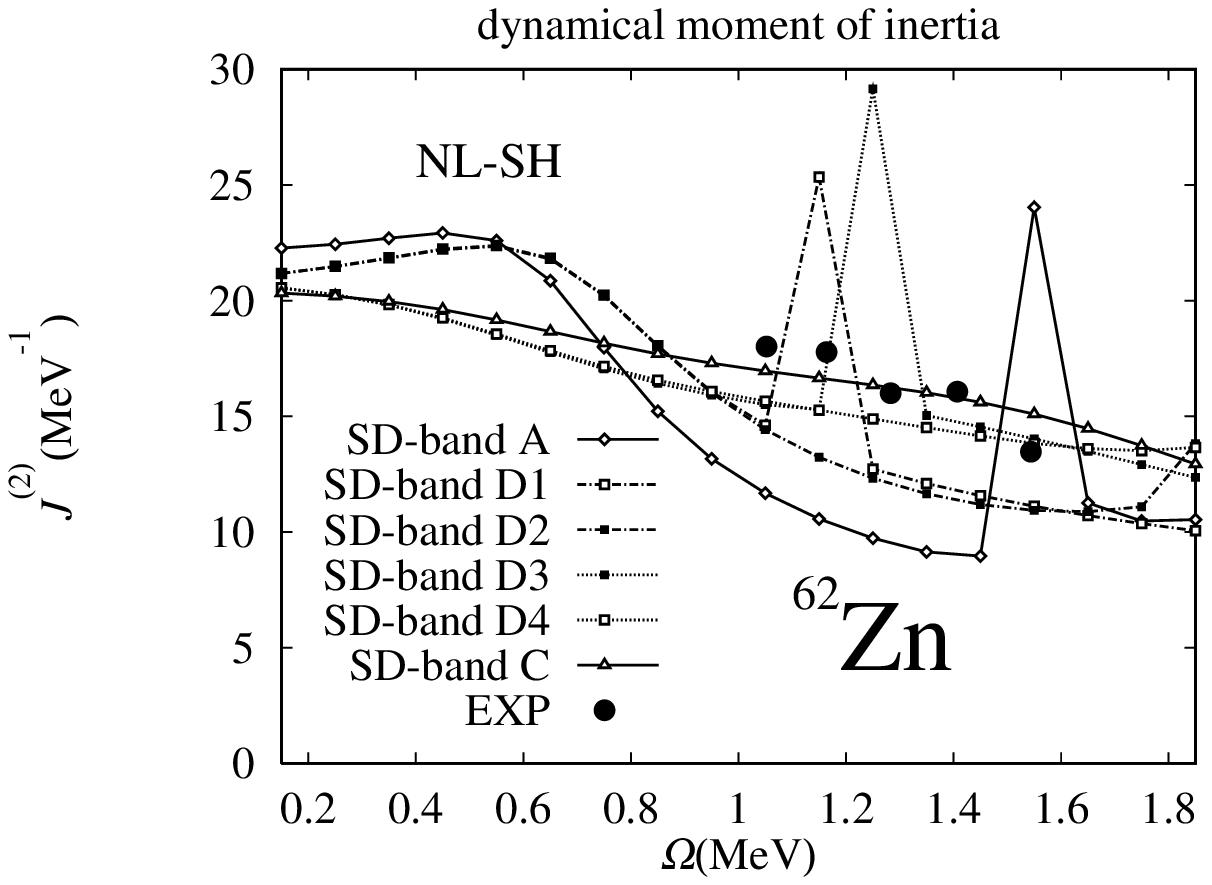}
  \epsfxsize=8cm
  \epsfbox{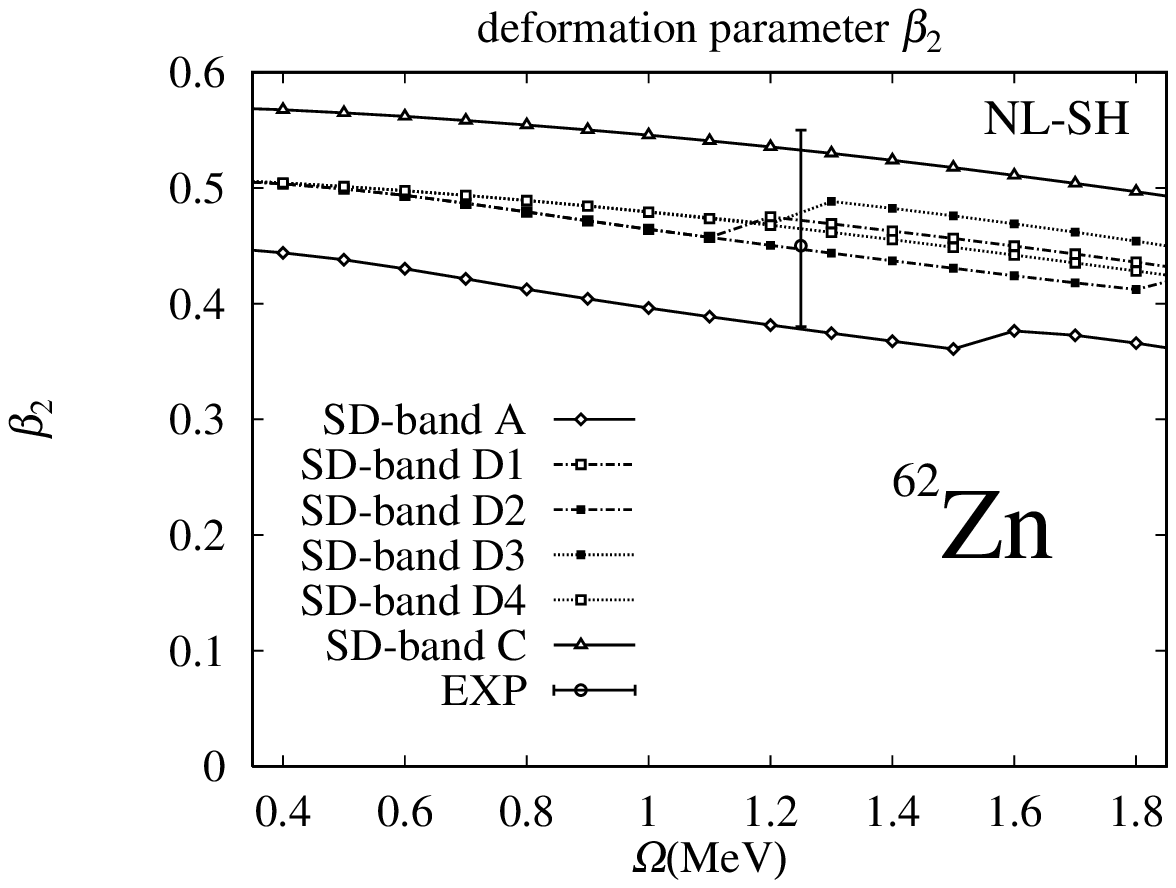}
  \epsfxsize=8cm
  \epsfbox{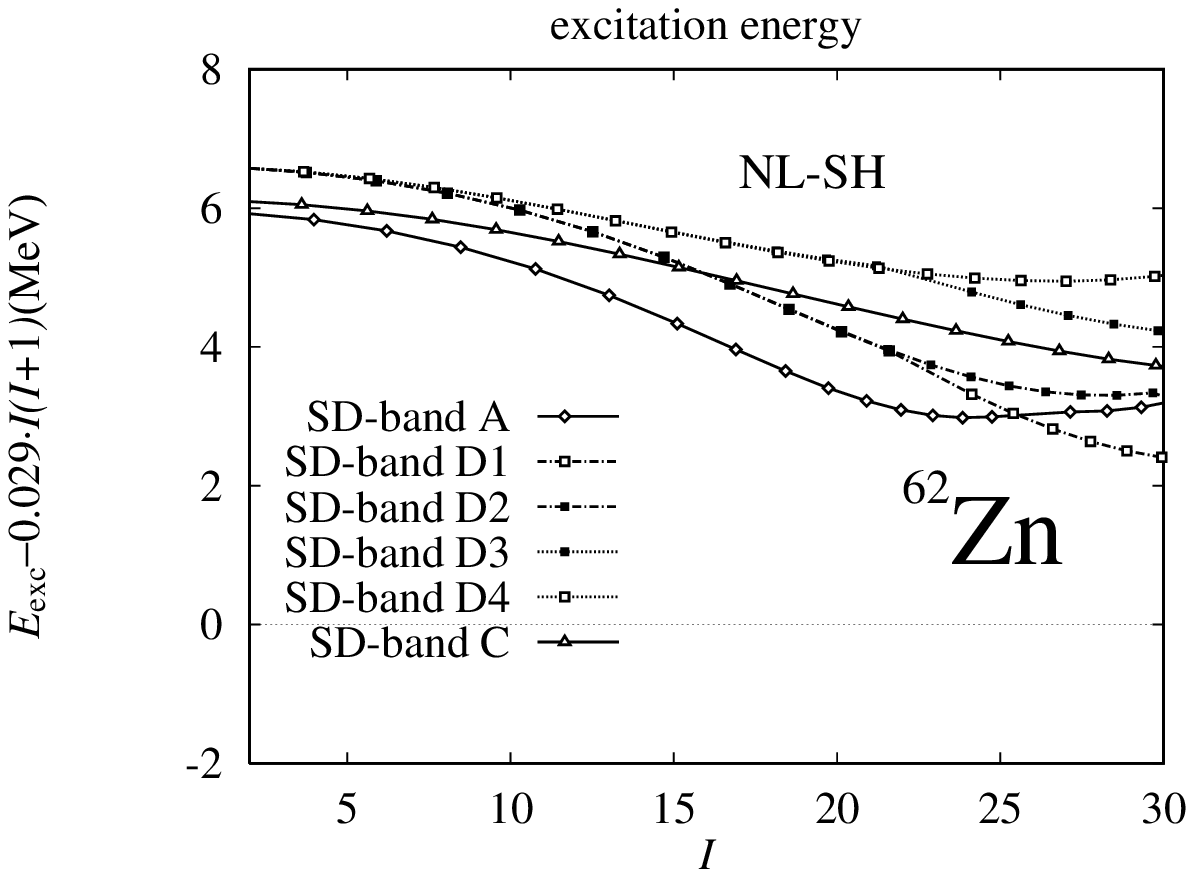}
  \caption{}
  \label{fig:62Zn_NL-SH}
\end{figure}

In the ground state of $^{62}$Zn, all valence nucleons are in the
$N_{\rm osc}$=3($pf$) orbits.  The SD states can then be generated by putting some
nucleons into the $N_{\rm osc}$=4 intruder orbits, and these configurations are
symbolically denoted as $\pi(3)^{-N_{\rm p}}(4)^{N_{\rm p}}
\nu(3)^{-N_{\rm n}}(4)^{N_{\rm n}}$.  According to ref.\cite{ref:Sv97},
the proton configurations are fixed to $N_{\rm p}$={\rm 2}.  Different SD
bands are then formed depending on the number of neutrons lifted into the
$N_{\rm osc}$=4 orbits, which is taken as $N_{\rm n}$=2--4 in this work.  We consider
the following SD configurations; A:$\pi(3)^{-2}(4)^{2}\nu(3)^{-2}(4)^{2}$,
D:$\pi(3)^{-2}(4)^{2}\nu(3)^{-3}(4)^{3}$ and
C:$\pi(3)^{-2}(4)^{2}\nu(3)^{-4}(4)^{4}$.  Here different configurations
are possible for D according to the parity-signature quantum number
($\pi=\pm,r=\pm i$) of the last nucleon. The occupation numbers of
each parity-signature block for these configurations are explicitly
written as \\
A\ :$\pi[8++;8+-;7-+;7--]\nu[8++;8+-;8-+;8--](\pi_{\rm tot}=+,
r_{\rm tot}=+1)$,\\
D1:$\pi[8++;8+-;7-+;7--]\nu[8++;9+-;8-+;7--](\pi_{\rm tot}=-,
r_{\rm tot}=+1)$,\\
D2:$\pi[8++;8+-;7-+;7--]\nu[8++;9+-;7-+;8--](\pi_{\rm tot}=-,
r_{\rm tot}=-1)$,\\
D3:$\pi[8++;8+-;7-+;7--]\nu[9++;8+-;8-+;7--](\pi_{\rm tot}=-,
r_{\rm tot}=-1)$,\\
D4:$\pi[8++;8+-;7-+;7--]\nu[9++;8+-;7-+;8--](\pi_{\rm tot}=-,
r_{\rm tot}=+1)$,\\
C\ :$\pi[8++;8+-;7-+;7--]\nu[9++;9+-;7-+;7--](\pi_{\rm tot}=+,
r_{\rm tot}=+1)$,\\
where the total parity and signature are also shown.  The calculated
dynamical moments of inertia of several SD bands in $^{62}$Zn are shown
in the top part of Fig.\ref{fig:62Zn_NL-SH}.  The experimental data are
also shown in the figure.  The sudden jumps in the bands A, D1 and D3 are
caused by the crossing in the single neutron levels between the
[303 7/2]($r=+i$) orbit and the [312 3/2]($r=+i$) one(see the upper part
of Fig.\ref{fig:snr62Zn}).  Apart from these crossings whose frequencies
depend on the choice of the parameter set, the bands D3, D4 and C
look to reproduce the experimental values well. The deformation parameters
$\beta_{2}$ are also calculated which are shown in the middle part of
Fig.\ref{fig:62Zn_NL-SH}.  The experimentally extracted value corresponding
to $I\sim$24 is shown with an error bar.  The bands D reproduce the
experimental value very well, while those of the bands A and C are somewhat
too large or too small.  From these results, we can consider that the
experimentally observed one corresponds to the bands D3 or D4
in our calculation.
The excitation energies of the bands mentioned above are shown in the bottom
part of Fig.\ref{fig:62Zn_NL-SH} relative to a rigid rotor reference.  For
$I\lsim$25 the band A is the lowest one while the band D1 comes down for
$I\gsim$25.  The bands D3 and D4 which give the best result are located more
than 1 MeV higher than the lowest one.  This result is consistent with the
calculation of the configuration-dependent shell correction approach shown
in refs.\cite{ref:Sv97,ref:Sv98}, but why only the band which
is not the lowest one is observed experimentally is not clear.

\begin{figure}[h]
  \epsfxsize=8cm
  \epsfbox{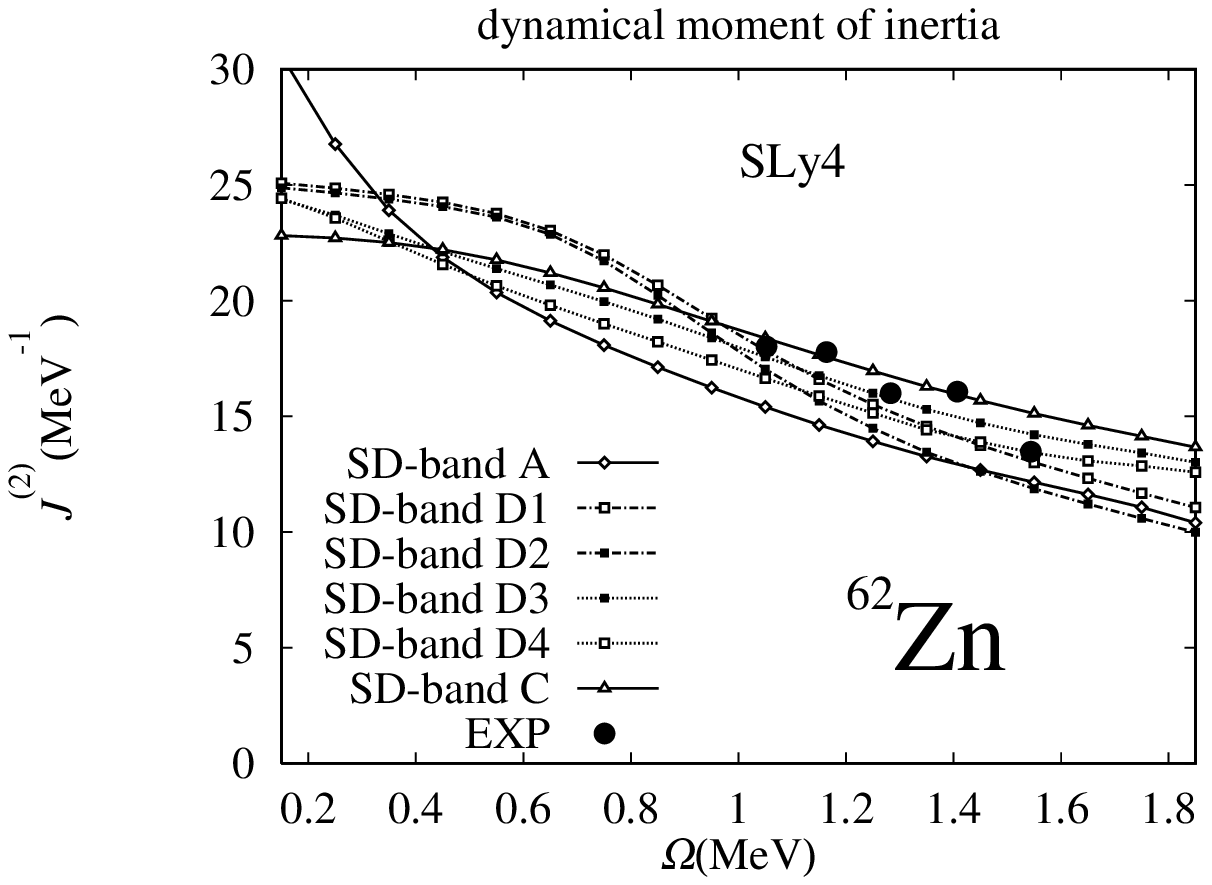}
  \epsfxsize=8cm
  \epsfbox{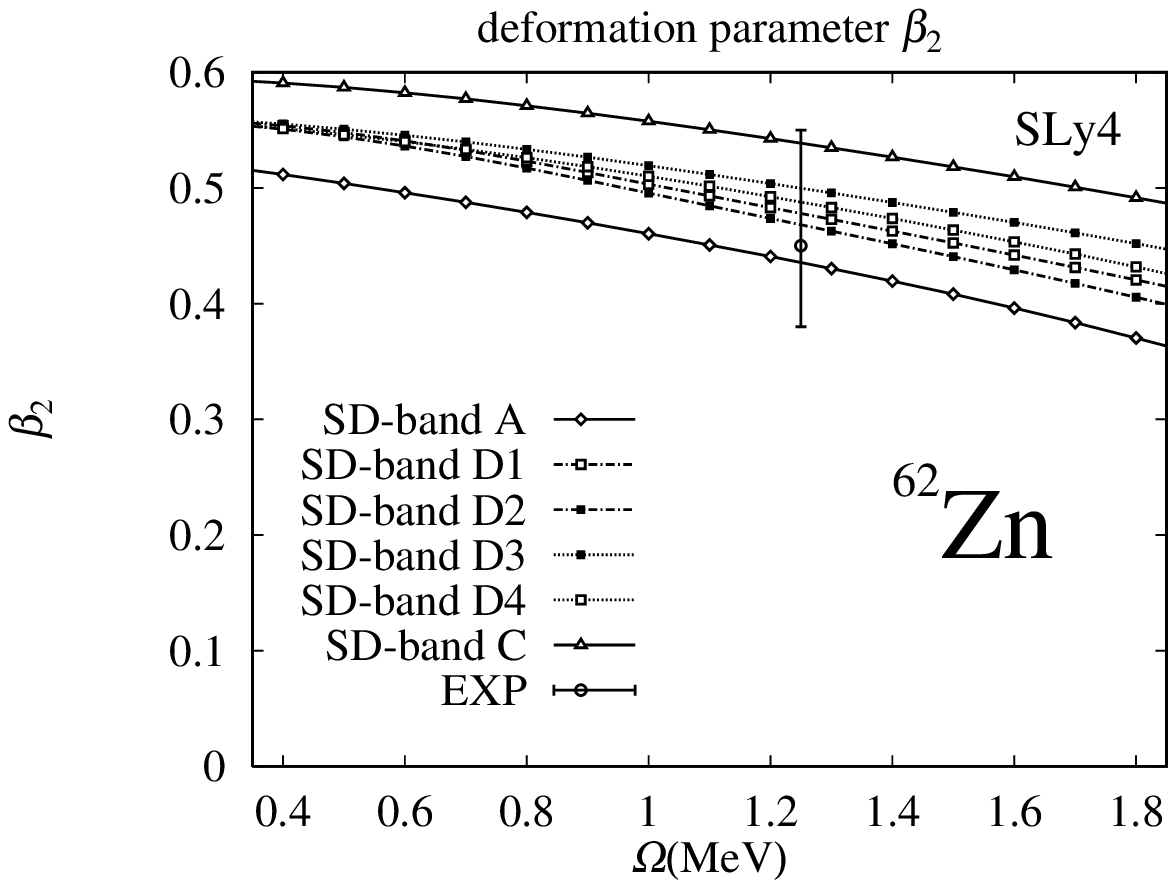}
  \epsfxsize=8cm
  \epsfbox{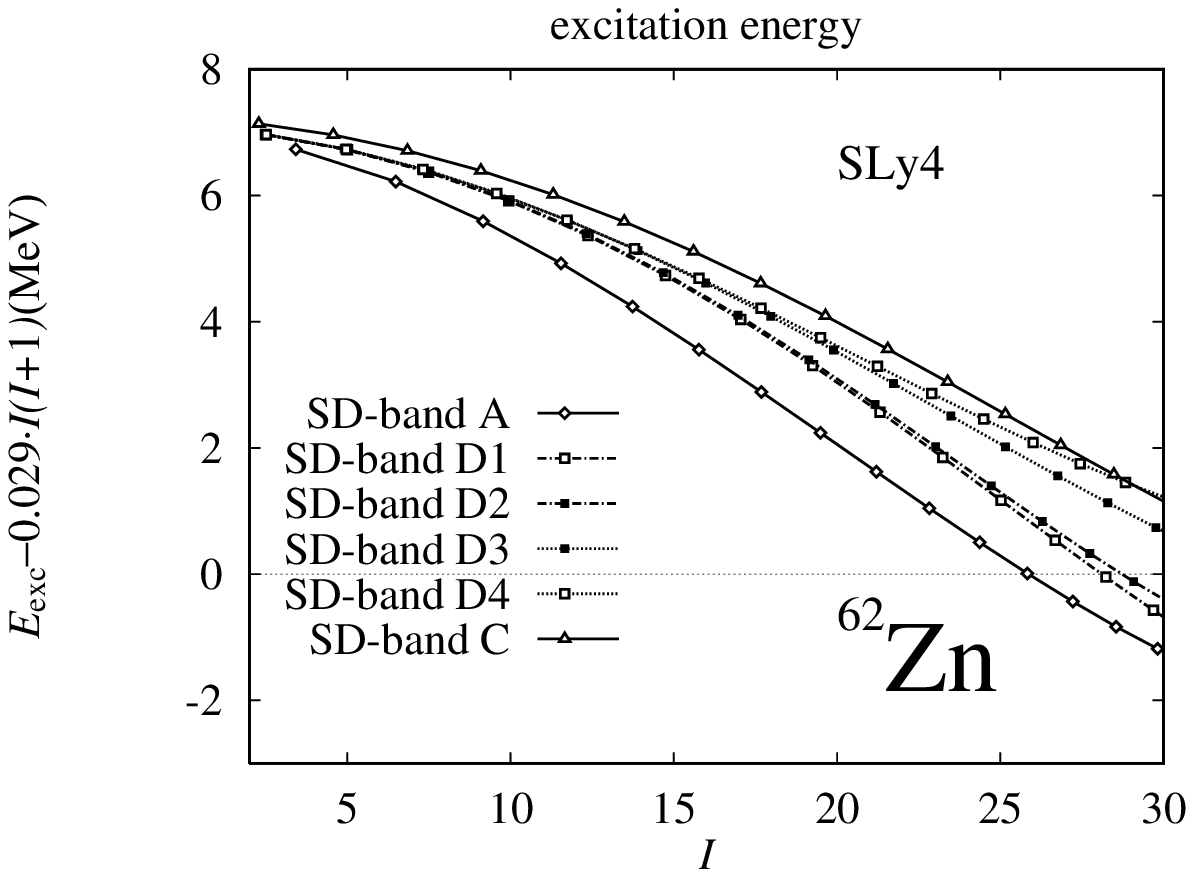}
  \caption{}
  \label{fig:62Zn_SLy4}
\end{figure}

\begin{figure}[h]
  \epsfxsize=12cm
  \epsfbox{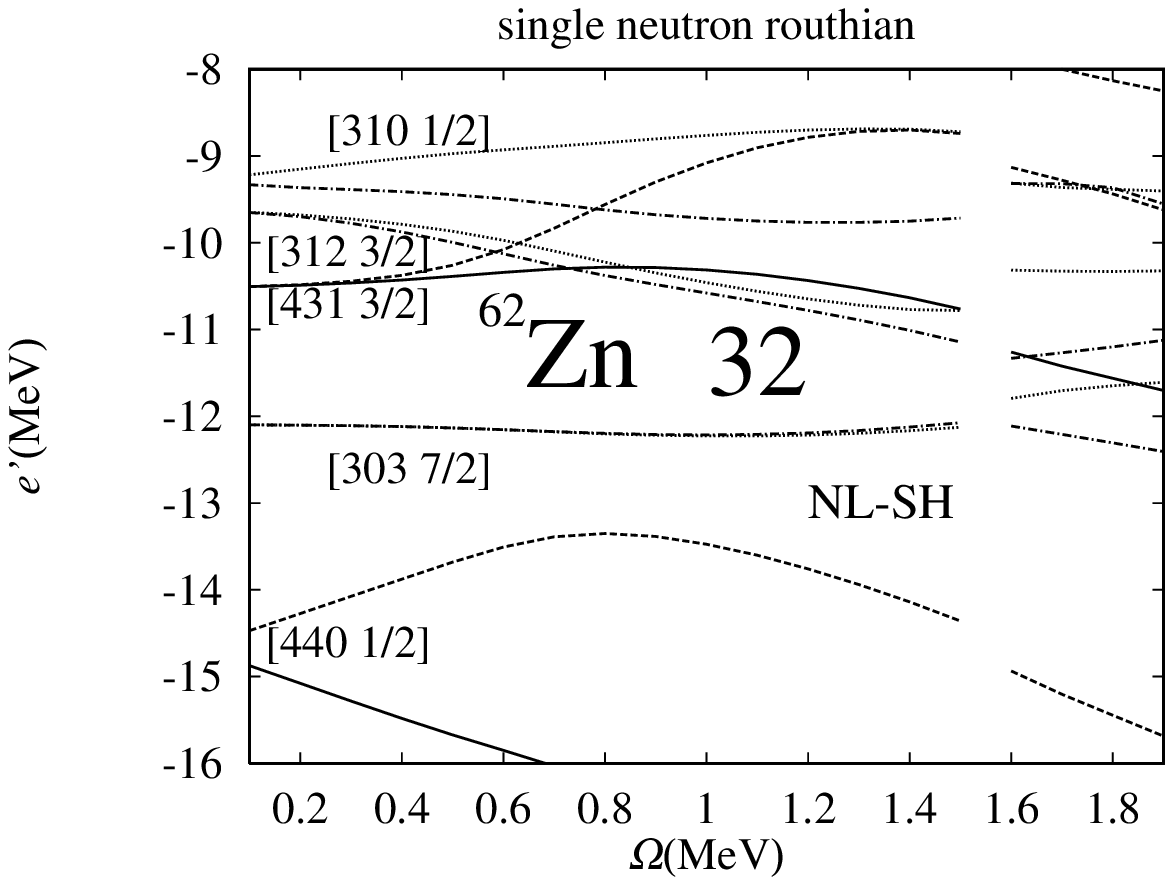}
  \epsfxsize=12cm
  \epsfbox{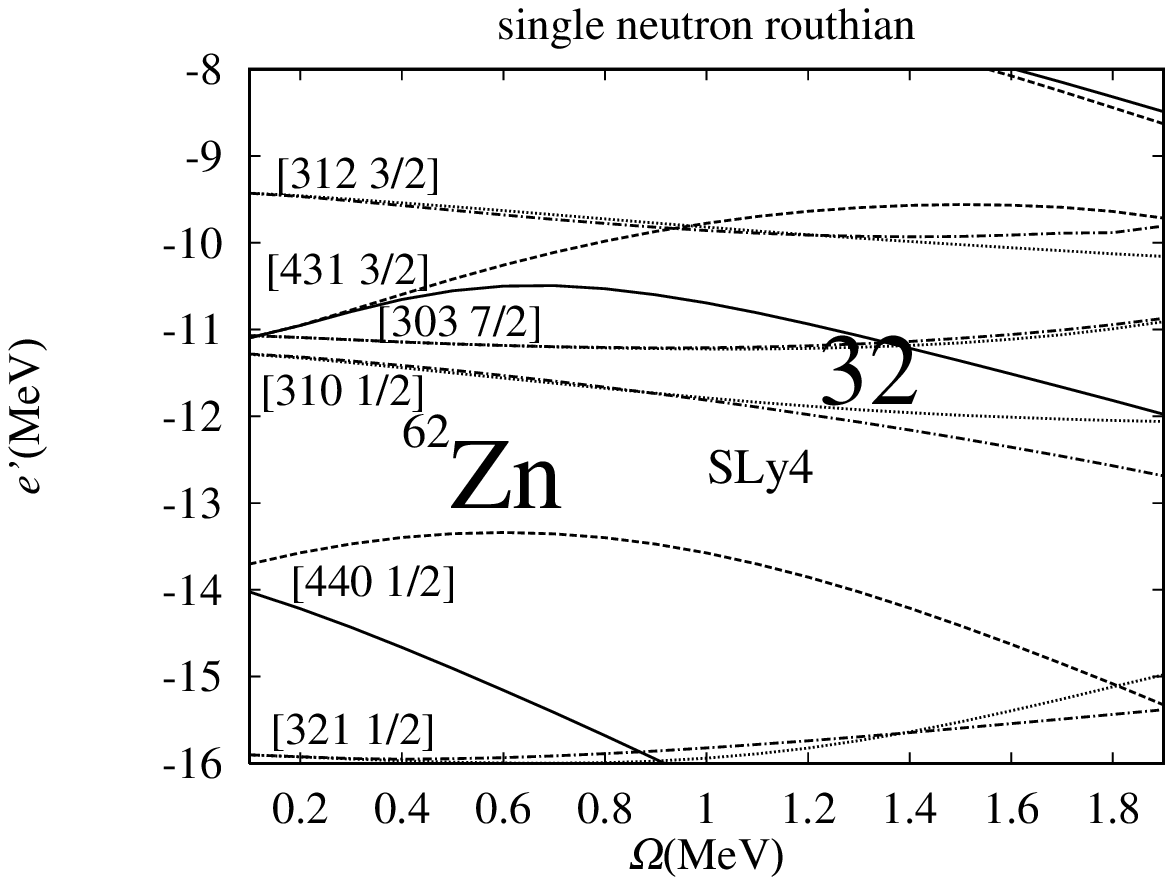}
  \caption{}
  \label{fig:snr62Zn}
\end{figure}

For comparison, we also perform SHF calculation of the SD bands in 
$^{62}$Zn, using the code HFODD which was developed by Dobaczewski and 
Dudek\cite{ref:DoDu97_1,ref:DoDu97_2,ref:Do98}.  The parameter set SLy4
\cite{ref:Ch95} is adopted since this set is more suitable for
$\beta$-unstable nuclei than other sets. Besides, other sets we examined
such as SkM* and SkP seem not to give the SD minima for $^{62}$Zn.  Figure
\ref{fig:62Zn_SLy4} shows the calculated moments of inertia, the deformation
parameters $\beta_{2}$ and the excitation energies of the same SD bands as
the RMF calculation in $^{62}$Zn.  For the band C, the moments of inertia was
already shown in Fig.3 of ref.\cite{ref:Ru98_2}.  Here we can see similar
trends to those of RMF shown in Fig.\ref{fig:62Zn_NL-SH}, that is, the bands
D3, D4 and C give the best results(although D1 is not so bad).
As for $\beta_{2}$, the bands D as well as the band A can reproduce
the experimental value.  The SHF calculation with SLy4 parameter set thus
also shows that the bands D3 and D4 are the best candidates for the
experimentally observed one.  Similarly to the RMF calculation,
these bands are again located more than 1 MeV higher than the lowest one,
which can be seen in the bottom part of Fig.\ref{fig:62Zn_SLy4}.
So far, no parity and signature assignment for these SD bands has been done.
If such assignment is completed and the experimentally observed band has
negative parity, we can determine which configuration corresponds to the
experimental one, because the band D4 has positive signature
($I^{\pi}$=0$^{-}$,2$^{-}$,4$^{-}$...) while the band D3 has negative
signature($I^{\pi}$=1$^{-}$,3$^{-}$, 5$^{-}$...).  On the other hand,
if the experimentally observed band has
positive parity, both models could not reproduce the experimental data,
which would imply that there are some other effects which are neglected in the
present calculation, such as residual (neutron-proton) pairing correlations
or the reflection asymmetric shapes in connection with the cluster-like
structure.

Looking at the results of the calculations closely, there exist some
qualitative differences between RMF and SHF, for example, the difference
of the deformation parameters, the occurrence of the crossing, and so on.
These differences are arising from those in the occupied single-particle 
orbits.  Figure \ref{fig:snr62Zn} shows the single neutron Routhians 
of the lowest SD band A as functions of the rotational frequency. 
All the orbits below the $N$=32 gap are occupied. We notice that the
position of the [310 1/2] orbit in the SHF calculation is quite different
from that in the RMF calculation.  It is located lower than the
[303 7/2] orbit, and therefore, the last two neutrons occupy the [310 1/2]
orbit in the SHF calculation rather than the [303 7/2] orbit in RMF,
which leads to the qualitative differences between RMF and SHF.
This difference in the single-particle levels can be directly related to 
that in the deformation parameters $\beta_{2}$ which can be seen in the middle
parts of Fig. \ref{fig:62Zn_NL-SH} and Fig. \ref{fig:62Zn_SLy4}.
The RMF calculation gives $\beta_{2}\sim$0.37(band A),0.49(band D3) and
0.46(band D4) at $\Omega\sim$1.3 MeV, while in the SHF calculation we obtain
$\beta_{2}\sim$0.43(band A), 0.50(band D3) and 0.48(band D4), respectively
(both calculations give rather small values of $\gamma$ deformation in all
of these SD bands).  These values are consistent with the calculations shown
in ref.\cite{ref:Sv97}, $\beta_{2}=0.41-0.49$, except for the band A in the RMF
calculation.  The reason why the deformations differ in these two models,
especially for the band A, is as follows: The [303 7/2] orbit is 
strongly upsloping with respect to $\beta_{2}$, {\it i.e.},
anti-deformation-driving, whereas the [310 1/2] orbit is almost flat
(Fig.1 of ref.\cite{ref:Na85}).  Because of this, the RMF calculation of the
band A where the [303 7/2] orbits are occupied gives smaller deformation than
the SHF calculation in which the [303 7/2] orbits are empty.  Both RMF and
SHF calculations of the SD bands in $^{60}$Zn, where the [303 7/2] orbits
in the single neutron routhian are not occupied, give very similar
deformations for the band A($\beta_{2}$=0.46 in both RMF and SHF at
$\Omega$=1.3 MeV) to those of ref.\cite{ref:Ra90}, $\beta_{2}$=0.47
($\epsilon_{2}$=0.41), which strongly supports these discussions given above.

\begin{figure}[h]
  \epsfxsize=16cm
  \epsfbox{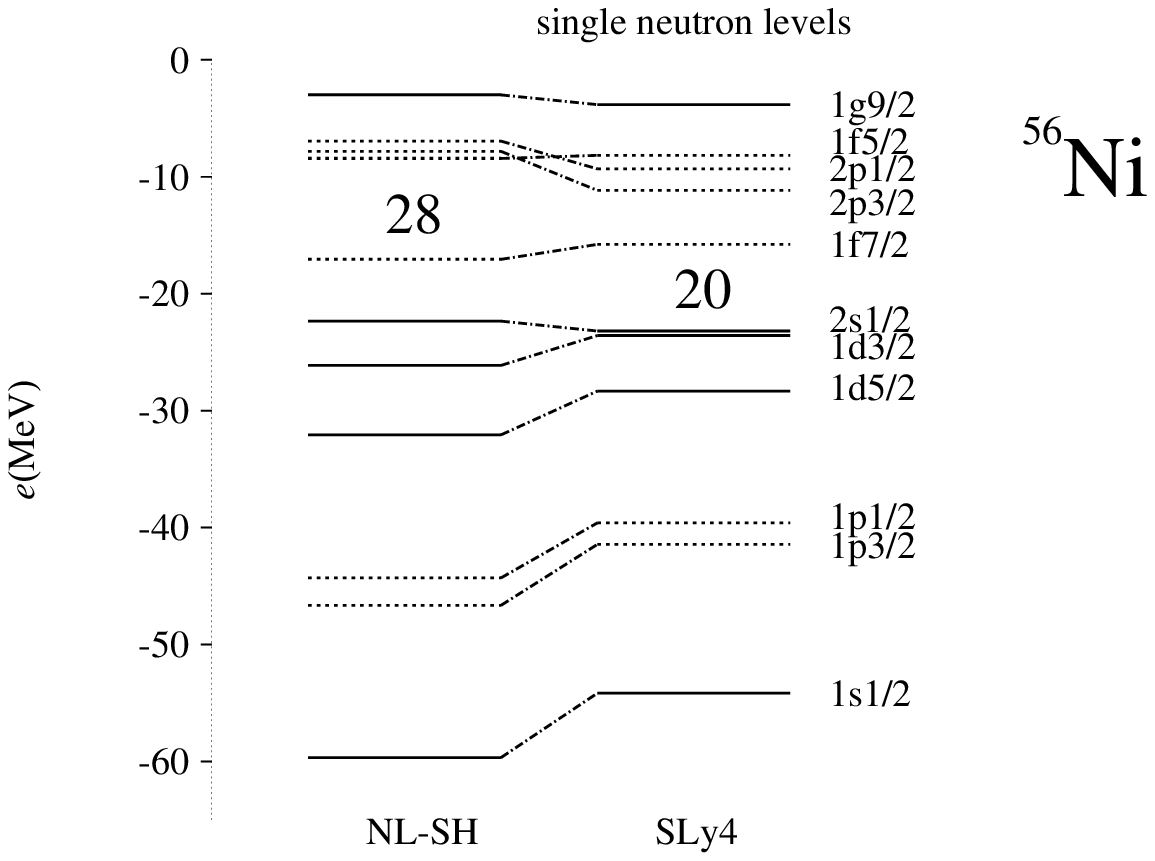}
  \caption{}
  \label{fig:sne56Ni}
\end{figure}

The relative position of the [310 1/2] orbit and the [303 7/2] one at zero
spin is determined by the spherical shell structure of both models.
To look into this, we calculate the single neutron levels in the spherical
core nucleus, $^{56}$Ni.  They are shown in Fig.\ref{fig:sne56Ni}.
Although the magnitudes of the $L$-$S$ splittings are very similar
in these two models, there are surely some differences, especially the
ordering in the $pf$ shells.  In the SHF calculation, the 2$p_{1/2}$ and 
2$p_{3/2}$ orbits are located below the 1$f_{5/2}$ orbit.  Because of this, 
the $N=$28 shell gap reduces and accordingly the [310 1/2] orbit is relatively 
lowered in SHF in comparison with the RMF calculation, which induces the 
different deformations and configurations in these two models for the SD 
bands in $^{62}$Zn.  

In summary, we studied the SD bands in $^{62}$Zn using RMF model,
including the recently discovered one.  For comparison, we also performed
SHF calculation of the SD bands in this nucleus.  In both the RMF calculation
with NL-SH parameter set and the SHF calculation with SLy4 parameter set,
the bands D3 and D4 give the best result.  There exist some qualitative
differences between the results of these two models.  They seem to be
arising directly from the difference in the equilibrium deformation, and it
can be understood qualitatively in terms of the difference in the order of
single-particle levels of the spherical core.  Because there is only little
experimental information up to now, it is too early to draw a definite
conclusion as for whether the present models can describe the very high spin
states in this mass region.  At the present stage, more systematic
investigation of many nuclei in this mass region using various parameter
sets of both RMF and SHF will be necessary, and which is under progress.

\ack
One of the authors(H.M.) would like to acknowledge the Junior Research
Associate Program of Japan Science and Technology Agency.
A part of numerical calculations in this work was done using the computer 
systems of Research Center for Nuclear Physics (RCNP), Osaka University.

\newpage

\noindent
Fig.\ref{fig:62Zn_NL-SH}: Moments of inertia, deformation parameters
and excitation energies of several SD bands in $^{62}$Zn calculated by
adopting the parameter set NL-SH.  Experimental values [3] are also shown in
the top and middle figures.

\noindent
Fig.\ref{fig:62Zn_SLy4}: Moments of inertia, deformation parameters
and excitation energies of several SD bands in $^{62}$Zn calculated by
adopting the parameter set SLy4.  Experimental values [3] are also shown in
the top and middle figures.

\noindent
Fig.\ref{fig:snr62Zn}: Single neutron Routhians of the SD shape in $^{62}$Zn
calculated by adopting the parameter sets NL-SH and SLy4. The discontinuity
in the upper figure is due to the crossing between the [303 7/2]($r=+i$)
orbit and the [312 3/2]($r=+i$) one.

\noindent
Fig.\ref{fig:sne56Ni}: Single neutron levels in the ground state of the doubly
magic $^{56}$Ni calculated by adopting the parameter sets NL-SH and SLy4.

\newpage

\newpage

\newpage

\newpage

\end{document}